\documentstyle[12pt]{article}

\title{Post-Post-Newtonian Limit of a Dilatonic Gravity Model.}

\author{Tonatiuh Matos\thanks{email:tmatos@fis.cinvestav.mx} \\
  Instituto de F\'{\i}sica y Matem\'aticas \\
  Universidad Michoacana de San Nicol\'as de Hidalgo \\
  A.P. 2-82, 58040 Morelia, Michoac\'an, M\'exico
\and Hugo Villegas-Brena\thanks{email:brena@fis.cinvestav.mx} \\
  Depto. de F\'{\i}sica, CINVESTAV IPN \\
  A.P. 14-740, 07000 M\'exico D.F., M\'exico }

\begin{document}

\maketitle

\begin{abstract}

We study a solution of the field equations for dilatonic gravity and
obtain its post-post-newtonian limit. It turns out that terms to this and
higher orders in the expansion may become important in strong
gravitational fields, even though the post-newtonian limit coincides with
that of General Relativity. This suggests that strong gravitational fields
can only be studied by exact solutions of the field equations.

\end{abstract}

In recent years there has been an increasing interest in tensor-scalar
theories of gravity, as they arise naturally in several unification
models. However, for the slow motion, weak field limit, the PPN formalism
~\cite{will} has proved to be a very powerful tool to restrict theories
that are phisically viable. On the other hand, with binary pulsar
observations the tests of the strong field regime have begun, but again,
the measurements have further limited the possible alternatives to General
Relativity~\cite{dam}. Here we will suggest, by means of an example, that
the PPN framework may be incomplete, and therefore, that exact solutions
to the field equations may be the only way to describe certain aspects of
physical systems in strong fields.

We shall be concerned with the action:

\begin{equation}
\label{eq:ac}
S=\int d^4x\sqrt{-g}(R-2(\nabla\phi)^2-e^{-2\alpha\phi}F^2),
\end{equation}

\noindent where $R$ is the scalar curvature, $\phi$ the dilaton,
$F_{\mu\nu}$ the Maxwell tensor and $\alpha$ the (non-minimal) coupling
parameter. For $\alpha=0,\sqrt{3},1$ we obtain back, respectively,
Einstein-Maxwell, Kaluza-Klein and Low-Energy Superstring theories, but we
shall consider the general case with arbitrary $\alpha$. It can be easily
checked that the metric~\cite{ton}:

\begin{equation}
\label{eq:metr}
ds^2=-(1-\frac{2m}{r})dt^2+e^{2k_s}\frac{dr^2}{1-\frac{2m}{r}}
+r^2(e^{2k_s}d\theta^2+\sin^2\theta d\phi^2),
\end{equation}

\noindent with

\[
e^{2k_s}=\left\{ 
 \begin{array}{cc}
   \left(1+\frac{m^2\sin^2\theta}{r^2(1-\frac{2m}{r})}\right)^{-1/\alpha^2}
       & \mbox{for $\alpha\neq 0$,} \\
   1   & \mbox{for $\alpha=0$ (Schwarzschild solution),}
 \end{array}
\right.
\]

\noindent and

\[
\Phi = \frac{1}{2\alpha}\ln\left(1-\frac{2m}{r}\right),
\]

\noindent is an axisymmetric static (indeed, quasi spherically symmetric)
solution of the field equations for the action (\ref{eq:ac}) in
Boyer-Lindquist coordinates.  The above expression for $e^{2k_s}$, which
is the contribution from the dilaton to the metric, can be rewritten as:

\[
e^{2k_s}=
\left(1+\frac{D^2\alpha^2\sin^2\theta}{r^2(1-\frac{2m}
{r})}\right)^{-1/\alpha^2},
\]

\noindent Here $D$ is the dilatonic charge, defined by ~\cite{dilch}:

\begin{equation}
\label{eq:dilch}
D=\lim_{r\rightarrow\infty}\oint_S
d^2S^\mu\nabla_\mu\phi=\frac{m}{\alpha}.
\end{equation}

Now, let us obtain the series expansion of the metric ~(\ref{eq:metr}) in
powers of $r$, in order to be able to compare it with the PPN expansions.
To do so, we perform a transformation to a ``pseudo-isotropic'' coordinate
system by means of: 

\[
r=R(1+\frac{m}{2R})^2,
\]

\noindent thus obtaining:

\newpage

\begin{eqnarray}
\nonumber ds^2 & = &
 e^{2k_s}(1+\frac{m}{2R})^2(dx^2+dy^2+dz^2)+
 \frac{(1-e^{2k_s})(1+\frac{m}{2R})^2}{x^2+y^2}(xdy-ydx)^2 \\
 & & \mbox{} -\frac{(1-\frac{m}{2R})^2}{(1+\frac{m}{2R})^2}dt^2,
\end{eqnarray}

\noindent where $(x,y,z)$ are related to $(R,\theta,\phi)$ through the
usual cartesian to spherical coordinate transformation. Note that, for
large $r$, $R\sim r$.  We have then the following expansions: 

\begin{eqnarray}
\nonumber g_{tt} & = &
 -(1-\frac{2m}{R}+\frac{2m^2}{R^2}-\frac{3m^3}{2R^3}+\cdots) \\
\nonumber g_{xx} & = & 
 1+\frac{2m}{R}+\frac{1}{R^2}\left[\frac{3m^2}{2}-D^2\frac{(x^2+y^2)}{R^2}+
 \frac{y^2}{R^2}\right]+\cdots \\
g_{yy} & = & 
 1+\frac{2m}{R}+\frac{1}{R^2}\left[\frac{3m^2}{2}-D^2\frac{(x^2+y^2)}{R^2}+
 \frac{x^2}{R^2}\right]+\cdots\\
\nonumber g_{zz} & = & 1+\frac{2m}{R}+\frac{1}{R^2}\left[\frac{3m^2}{2}-
 D^2\frac{(x^2+y^2)}{R^2}\right] +\cdots\\
\nonumber g_{xy} & = & -\frac{1}{R^2}\left(\frac{xy}{R^2}D^2\right)+\cdots
\end{eqnarray}

The post-newtonian limit corresponds to terms $O(1/R^2)$ in $g_{tt}$ and
$O(1/R)$ in $g_{ij}$, and it coincides with the same limit for General
Relativity. It should not be surprising then if this metric passes the
classical tests for the solar system, that is, for a weak field, slowly
moving system of particles (for a study of this metric in the solar system
see \cite{ma97}), even when the scalar interaction is taken into account.
Observe that the dilatonic charge $D$ appears in the expansion only at
orders of $O(1/R^2)$ in $g_{ij}$, which correspond to the
post-post-newtonian expansion. Surprisingly, this charge cannot be
eliminated from the metric for $\alpha\neq 0$ if the mass parameter $m\neq
0$. But, contrary to what happens in the PPN expansion with scalar field,
here the interaction of the scalar field in the metric is weaker as the
coupling parameter $\alpha$ gets larger. This is because $D$ has a
discontinuity for $\alpha=0$, as can be seen from (\ref{eq:dilch}): 

\[
\lim_{\alpha\to 0}{D}=\infty\neq D\mid_{\alpha=0}=0.
\]

For the post-post-newtonian limit, corresponding to terms $O(1/R^4)$ in
$g_{tt}$ and $O(1/R^2)$ in $g_{ij}$, $\alpha$ appears in the metric terms
through the value of $D$. Again contrary to the PPN expansion of scalar
theories of gravity, the predictions of this theory will be closer to
those of General Relativity if the value of $\alpha$ is large, that is, if
the coupling with the scalar field is big. Of course, the special case
$\alpha=0$ should be treated just as the Schwarzschild solution.

Scalar theories of gravity have been studied using either spherically
symmetric exact solutions, or the PPN formalism. This analysis shows that
maybe it was too premature to discard some of these theories using only
the former methods, and thus, assuming that the metric (\ref{eq:metr}) 
could represent an actual physical system, it should be possible to
describe effects that would be otherwise overlooked using the PPN
formalism, at least for certain values of $\alpha$. Finally, it remains to
find out whether this metric actually passes the standard tests for the
solar system, and whether its predictions for strong fields agree
with binary pulsar observations.


\begin{thebibliography}{99}

\bibitem{will} C.M. Will,
``{\it Theory and Experiment in Gravitational Physics}'',
Cambridge University Press (1981).

\bibitem{dam} T. Damour and G. Esposito-Far\`ese,
{\it Phys. Rev.} {\bf D54}, 1474 (1996).

\bibitem{ton} T. Matos, D. Nu\~nez and H. Quevedo, 
{\it Phys. Rev.} {\bf D51}, R310 (1995).

\bibitem{dilch} D. Garfinkle, G. T. Horowitz and A. Strominger,
{\it Phys. Rev.} {\bf D43}, 3140 (1991).

\bibitem{ma97} T. Matos, {\it Can Magnetic Fields of Astrophysical Objects be Fundamental?}. To be Published. T. Matos and C. Mora, 
{\it Class. Quantum Grav.} {\bf 14}, in Press.



\end{thebibliography}
\end{document}